%% file: main.tex
\documentclass[10pt,conference]{IEEEtran}
\usepackage[T1]{fontenc}
\usepackage{cite}
\usepackage{amsmath,amssymb,amsfonts}
\usepackage{algorithmic}
\usepackage{graphicx}
\usepackage{textcomp}
\usepackage{xcolor}
\usepackage{url}
\usepackage{tabularx}
\usepackage{booktabs}
\usepackage{array}

\begin{document}

\title{Req-Space: A User Study of Interactive Visualisations for Requirements Review}

\author{
\IEEEauthorblockN{Risha Parveen}
\IEEEauthorblockA{
\textit{Tampere University}\\
Tampere, Finland\\
risha.parveen@tuni.fi\\
ORCID: 0009-0009-4515-8788
}
\and
\IEEEauthorblockN{Navneet Arora}
\IEEEauthorblockA{
\textit{Tampere University}\\
Tampere, Finland\\
navneet.arora@tuni.fi\\
ORCID: 0000-0002-7329-8261
}
\and
\IEEEauthorblockN{Zheying Zhang}
\IEEEauthorblockA{
\textit{Tampere University}\\
Tampere, Finland\\
zheying.zhang@tuni.fi\\
ORCID: 0000-0002-6205-4210
}
\and[\hfill\par\vspace{1em}\hfill]
\IEEEauthorblockN{Kalle Mäkelä}
\IEEEauthorblockA{
\textit{Aimbition Oy}\\
Tampere, Finland\\
kalle.makela@aimbition.com\\
ORCID: 0009-0004-8602-2952
}
\and
\IEEEauthorblockN{Pekka Abrahamsson}
\IEEEauthorblockA{
\textit{Tampere University}\\
Tampere, Finland\\
pekka.abrahamsson@tuni.fi\\
ORCID: 0000-0002-4360-2226
}
}

\maketitle

\

\begin{abstract}
Requirements review involves identifying inconsistencies and assessing change implications across distributed specifications. Motivated by needs raised in industrial project workshops, we investigate Req-Space, an interactive prototype integrating semantic exploration and traceability visualisation to support human-led review. We conducted a fixed-order, two-phase study with 20 participants, including 11 with at least three years of software or IT experience. Participants reviewed a fictional 40-requirement software specification and a guest-checkout change request, first using source documents and then revisiting the same tasks with Req-Space. During the supported review, 17 participants recorded at least one additional source-supported item or pair. Of 69 assessable additional units, 58 (84.1\%) were judged supported by the source materials. Seven decision rationales were judged stronger in Phase~2, including five without a change in the decision category.  Participants generally found the visualisations helpful, although workload responses were mixed and some misinterpreted visual relationships. The findings suggest that visual support can facilitate requirements exploration and decision justification, provided relationship meanings and source context remain clear. The fixed study sequence prevents attributing the observed changes specifically to Req-Space.
\end{abstract}


\input{sections/introduction}

\input{sections/background}

\input{sections/reqspace}

\input{sections/study_design}

\input{sections/results}

\input{sections/discussion}

\input{sections/threats_to_validity}

\section{Conclusion}
\label{sec:conclusion}
We investigated Req-Space for reviewing requirements and proposed changes, distinguishing broader review coverage from stronger justification of existing decisions. Participants valued access to related requirements and traceability information, but correct interpretation and source inspection remained essential. The findings suggest a role for visualisations in supporting human-led review, although the fixed-order design prevents causal attribution of benefits to Req-Space. Its use in collaborative software development remains to be evaluated.

\section*{Data Availability}

The Req-Space study application, study materials,
questionnaires, and analysis scripts will be available at \url{https://github.com/risha-parveen/req-space-study}. The replication package and anonymised research data supporting the results will be archived on Figshare(\url{https://doi.org/10.6084/m9.figshare.34767168}). Raw participant data that cannot be safely anonymised will not be publicly shared to protect participant privacy.


\section*{Acknowledgments}
This work was supported by Business Finland through the ANSE project.

\textit{Declaration of AI-Assisted Technologies.}
ChatGPT (including GPT-5.6) assisted with fictional study materials, manuscript drafting and editing, and qualitative code grouping; Codex assisted with application and script development. The authors reviewed and approved the final coding, interpretations, and manuscript and take full responsibility for the work.

\bibliographystyle{IEEEtran}
\bibliography{refs}

\end{document}

%% file: sections/introduction.tex
\section{Introduction}
\label{sec:introduction}


Reviewing software requirements is essential for identifying inconsistencies, clarifying ambiguities, and assessing the implications of proposed changes~\cite[Ch.~1, Sec.~5.1]{swebok}. In industrial software projects, these activities are complicated by requirements distributed across specifications, features, and teams~\cite{reddivari2014visual}. Requirements concerning the same functionality may appear in different parts of a specification, while a proposed change may affect requirements, business rules, and goals beyond those explicitly linked to the modified item. The Naming the Pain in Requirements Engineering (NaPiRE) study~\cite{namingthepain} similarly identifies inconsistent and changing requirements among recurrent problems in practice.

The motivation for this work also emerged from the Business Finland-funded AI-Native Software Engineering (ANSE) project~\cite{ANSE_project}. During project workshops, participating companies highlighted difficulties in understanding requirements relationships and assessing change implications. These discussions motivated our investigation of visual support that helps engineers navigate requirements evidence and make informed review decisions.

Requirements visualisations can support this process by making related information easier to locate, navigate, and compare. Existing approaches include similarity-based exploration of requirements~\cite{reddivari2012recvisu} and graph-based navigation of documented relationships~\cite{heim2008graph}. However, semantic similarity does not establish inconsistency, and traceability links do not necessarily identify all consequences of a change~\cite{clelandhuang2014traceability}. Reviewers must therefore interpret visual evidence in the context of the original requirements.



To address these needs, we developed Req-Space, an interactive prototype for requirements review. Its \emph{Requirements map} supports exploration of semantically related wording and side-by-side comparison, whereas \emph{Change and traceability} presents a proposed change and recorded relationships among requirements, business rules and goals. Both retain access to source documents and leave substantive judgements to reviewers. 

We evaluated Req-Space in a fixed-order, two-phase study using QuickBite, a fictional food-delivery system. The study involves 20 participants with varying IT industrial backgrounds. Participants first reviewed a fictional 40-requirement specification and a proposed change using source documents, then revisited the tasks with Req-Space. We analysed review records, interaction logs, and questionnaire responses to address three research questions: 

\begin{description}
    \item[\textbf{RQ1:}] What additions and revisions do reviewers make to their review records when revisiting requirements with Req-Space?
    \item[\textbf{RQ2:}] Which Req-Space features do reviewers use, and how helpful do they find them for requirements review and change-impact assessment?
    \item[\textbf{RQ3:}] What benefits, difficulties, and further support needs do reviewers report when using Req-Space?
\end{description}

This study contributes an industry-motivated approach to visual requirements review, empirical evidence on how reviewers extend and reconsider their assessments with interactive support, and practical design lessons for integrating semantic exploration and traceability into review activities. The findings indicate opportunities to broaden review coverage and strengthen decision justifications, while revealing difficulties in interpreting visual relationships. Although the fixed-order design does not establish an independent causal effect of Req-Space, the results offer insights for developing human-centred requirements review support in software engineering practice.

%% file: sections/background.tex
\section{Background and Related Work}
\label{sec:background}

\subsection{Requirements Review in Practice}
Requirements review addresses recurrent problems reported by practitioners. The NaPiRE study collected evidence from 228 companies in ten countries and identified problems including inconsistent requirements and moving targets, such as changes to requirements and goals~\cite{namingthepain}. A broader, cross-industry perspective comes from the Project Management Institute's 2018 survey: 35\% of respondents selected inaccurate requirements gathering as one of up to three primary causes of project failure in their organisations, while respondents estimated that 52\% of projects completed in the preceding year had experienced scope creep or uncontrolled changes to scope~\cite[p.~25]{pmi2018pulse}. These survey findings indicate practical concerns; they do not measure the effectiveness of requirements reviews or visualisation tools.

Established requirements engineering guidance emphasises consistency, traceability, and change management. The IREB CPRE Foundation Level syllabus identifies consistency, non-redundancy, and traceability as requirements quality criteria 
and covers requirements dependencies, change management, and task-appropriate tool support~\cite[Secs.~3.8, 6.6--6.7, 7.1]{ireb2026cpre}. In a domain-specific example, Automotive SPICE 4.0 addresses requirements interdependencies, consistency, and bidirectional traceability, as well as the assessment of change requests and affected work products~\cite{vda2023aspice40}. Its guidance distinguishes establishing trace links from demonstrating consistency. ISO/IEC/IEEE 29148:2018 defines requirements traceability in terms of documented derivation and allocation paths, linking requirements to higher-level needs and lower-level
implementation~\cite{iso29148_2018}.

These motivate the review activities considered here: gaining an overview of a specification, comparing statements that may be inconsistent, and investigating possible consequences of a proposed change. Each activity requires more than locating similar wording or finding a link. The reviewer may need to examine requirements alongside rules, goals and definitions before recording a justified judgement. Our study concerns these review tasks, not the organisational causes of requirement changes or their implementation.

\subsection{Visual Support and Evaluation}

Requirements visualisation uses visual representations of requirements information and relationships to support understanding and requirements-engineering activities~\cite{cooper2009requirements,abad2016requirements}. These include tabular, hierarchical, relational and quantitative views that complement textual specifications. Cooper et al.~\cite{cooper2009requirements} classify approaches by representation and supported activity, while Abad et al.~\cite{abad2016requirements} identify usage patterns and uneven activity coverage, including verification. Both emphasise alignment between representations and tasks. Shneiderman's information-seeking framework motivates overview, filtering and details on demand~\cite{shneiderman1996eyes}, while Munzner's Nested Model distinguishes domain tasks, data abstractions, visual encodings and algorithms~\cite{munzner2009nested}.

\emph{Similarity-based approaches} help reviewers discover potentially related requirements. Ferrari et al.~\cite{ferrari2012flaws} use lexical and syntactic clustering to identify departures from surrounding patterns, particularly incomplete descriptions and inconsistent terminology rather than general logical contradictions. ReCVisu~\cite{reddivari2012recvisu} combines textual similarity, layout-based clustering, and interaction to explore requirements themes and relationships, with initial feedback from four professional analysts. Reddivari et al.~\cite{reddivari2014visual} subsequently developed a framework connecting users, data, models, visualisations and knowledge, investigating exploration and decision support in an industrial case study. These studies support visual candidate discovery and task-oriented exploration without treating proximity as evidence of inconsistency.

Sentence-BERT provides sentence embeddings for cosine-similarity retrieval~\cite{reimers2019sentencebert}, while t-SNE projects high-dimensional representations into a low-dimensional map that preserves aspects of local neighbourhood structure~\cite{vanDerMaaten2008tsne}. However, similar wording may express compatible constraints, and projections cannot preserve all relationships in the original space. Similarity and position therefore provide candidate comparisons whose consistency requires inspection of the underlying statements.

\emph{Traceability visualisation} provides complementary support for investigating documented relationships. Heim et al.~\cite{heim2008graph} identify limitations of lists, trees and matrices for exploring multiple relationship types and propose graph-based focus-and-context exploration. Their prototype was not comparatively evaluated. Cleland-Huang et al.~\cite{clelandhuang2014traceability} highlight the effort needed to maintain trustworthy trace links and provide task-driven visual access: matrices show trace coverage, while graphs support navigation. In a controlled maintenance study with 52 students, M\"ader and Egyed~\cite{mader2012traceability} found faster and more correct solutions with requirements-to-code traces. This supports traceability during maintenance, but does not establish the effectiveness of requirements-to-requirements graphs for review.

These studies distinguish finding candidates, following
recorded relationships and judging their implications, which
requires inspecting statements and context. Empirical validation
and comparative evaluation remain limited
in requirements visual analytics~\cite{varikuti2024visual}.
Specifically, evidence remains limited on how reviewers use
complementary semantic and traceability views when revisiting
an assessment, what findings they add or revise, and what
interpretation difficulties arise.

Our study examines these questions using Req-Space, which combines semantic exploration with documented traceability. We investigate how reviewers extend and reconsider review records, justify change decisions, and identify limitations of visual support, rather than proposing automated inconsistency detection or change-impact assessment.

%% file: sections/reqspace.tex
\section{Req-Space: Design Rationale and Visualisations}
\label{sec:reqspace}

\subsection{Purpose and Design Rationale}

Req-Space is an interactive prototype supporting two requirements-review activities: identifying potentially inconsistent requirements and investigating proposed changes. It combines a semantic Requirements map with a Change and traceability workspace, providing complementary ways to explore evidence while retaining access to the original specification.

Three principles guided its design.
\emph{First, task-oriented representations:}
the Requirements map supports discovery and comparison of potentially related statements, while Change and traceability supports investigation of recorded SRS relationships associated with a proposed change~\cite{munzner2009nested,reddivari2014visual}.
\emph{Second, progressive exploration:}
overview, selection, filtering, and details on demand help reviewers focus on relevant information~\cite{shneiderman1996eyes}, while selective graph expansion limits visual clutter~\cite{heim2008graph}.
\emph{Third, evidence-based judgement:}
visual relationships guide investigation, while reviewers inspect requirements, business rules, and goals before recording findings or decisions. The tool neither generates inconsistency verdicts nor predicts change impacts.

\begin{figure*}[t]
    \centering
    \includegraphics[width=0.78\textwidth]{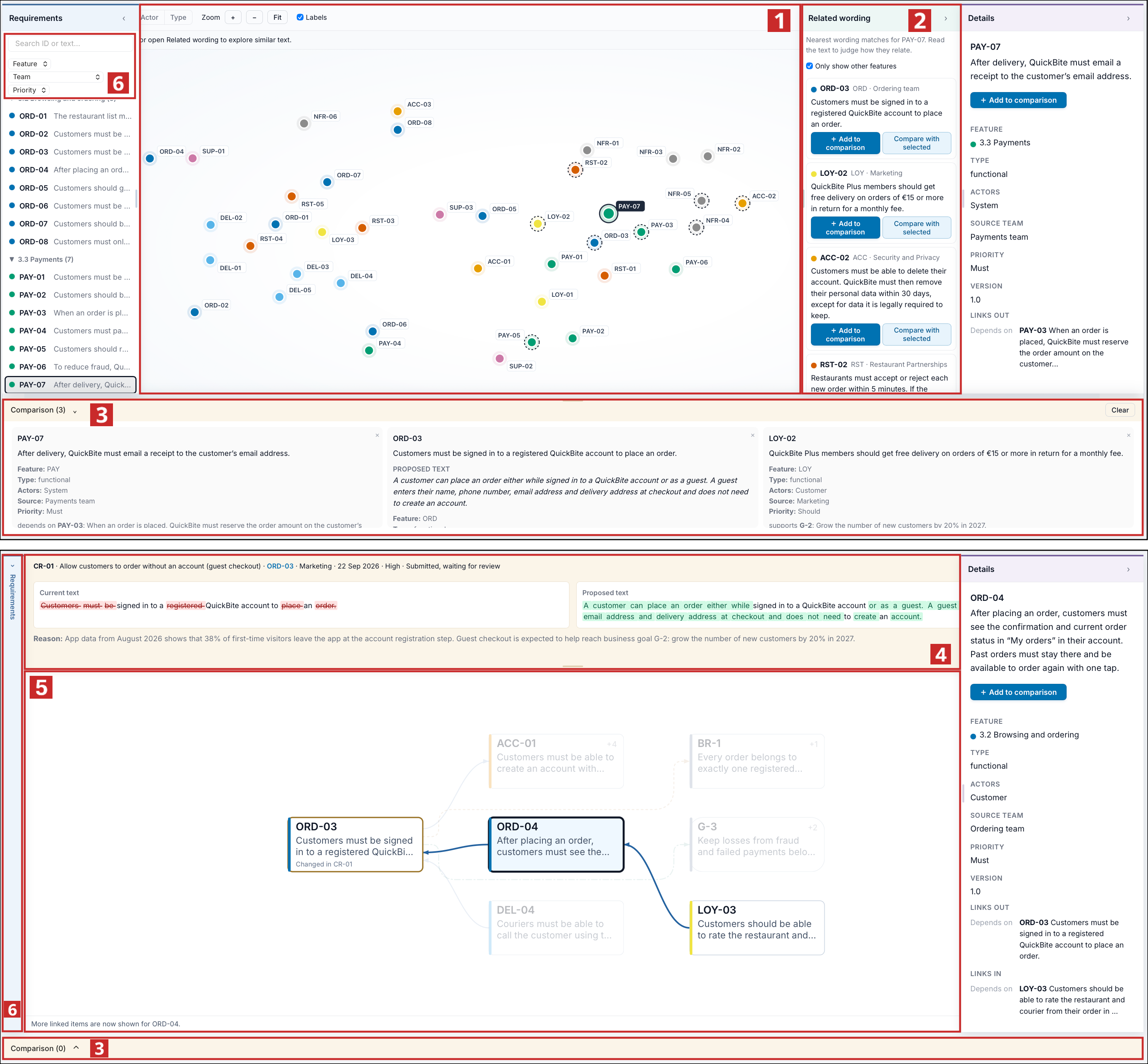}
    \caption{Req-Space workspaces in the evaluated build. Top: the Requirements map (1),
        related-wording panel (2), comparison tray (3) and requirements navigation (6).
        Bottom: the proposed-change panel (4), expandable relationship diagram (5), and
        shared comparison tray (3). Item details remain available alongside both views.}
    \label{fig:reqspace-interface}
\end{figure*}

The authors iteratively refined candidate representations and review tasks through informal discussions before selecting the two workspaces shown in Fig.~\ref{fig:reqspace-interface}. The evaluated build remained fixed across participants, with semantic representations and layouts prepared in advance and no runtime language-model assessments.

\subsection{Requirements Map}

Following clustering-based requirements exploration~\cite{ferrari2012flaws,reddivari2012recvisu}, the requirements map displays the 40 QuickBite requirements as labelled dots to help reviewers locate statements worth comparing across features and source teams.

Deterministic parsing converted the SRS into structured records, preserving wording, identifiers, source teams and documented relationships; fixed rules derived actor labels from requirement wording. Before the study, requirement text, excluding identifiers and metadata, was encoded using the pretrained \texttt{all-mpnet-base-v2} sentence-transformer model for cosine-similarity comparisons~\cite{reimers2019sentencebert}. A fixed t-SNE projection determines dot positions~\cite{vanDerMaaten2008tsne}. Because projected distances can distort semantic relationships, the separate Related wording panel retrieves eight nearest neighbours by cosine similarity in the original embedding space. It defaults to neighbours from other features, with an option to remove this filter. Neither view displays similarity scores or conflict judgements.

Reviewers can colour dots by feature, source team, actor or requirement type; focus legend categories; search the requirement list; and zoom or pan while preserving relative positions. Selecting a requirement reveals its full wording, attributes and recorded connections. The shared comparison tray displays up to four selected items side by side, with the SRS available for context. These controls support discovery and assessment, but relevant pairs may be missed and proximity does not establish a conflict.

\subsection{Change and Traceability}

The Change and traceability workspace presents CR-01's proposed replacement of mandatory sign-in with guest checkout, its rationale, and the current and proposed wording. Word-level highlighting identifies textual changes, with the original change request available for context.

A directed, typed graph shows recorded SRS relationships among requirements, business rules and goals. It initially displays the requirement targeted by CR-01 and its immediate incoming and outgoing links. Types include \emph{depends on}, \emph{refines}, \emph{follows} a business rule and \emph{supports} a business goal. Selecting a card reveals further directly connected items; reviewers can filter relationship types and inspect full wording. Progressive expansion limits the visual clutter associated with dense graphs~\cite{heim2008graph}.

Recorded links may be incomplete, and connected items may require no revision~\cite{clelandhuang2014traceability}. Reviewers can therefore add any SRS requirement, rule or goal, including items without a recorded path to the changed requirement. Added items remain visibly distinct; the tool neither infers links nor labels them as affected. Reviewers can inspect their full text in the detail panel, compare them alongside the proposed wording in the shared tray, and consult the original documents.

%% file: sections/study_design.tex
\section{Study Design}
\label{sec:study-design}

\begin{figure}
    \centering
    \includegraphics[width=1\linewidth]{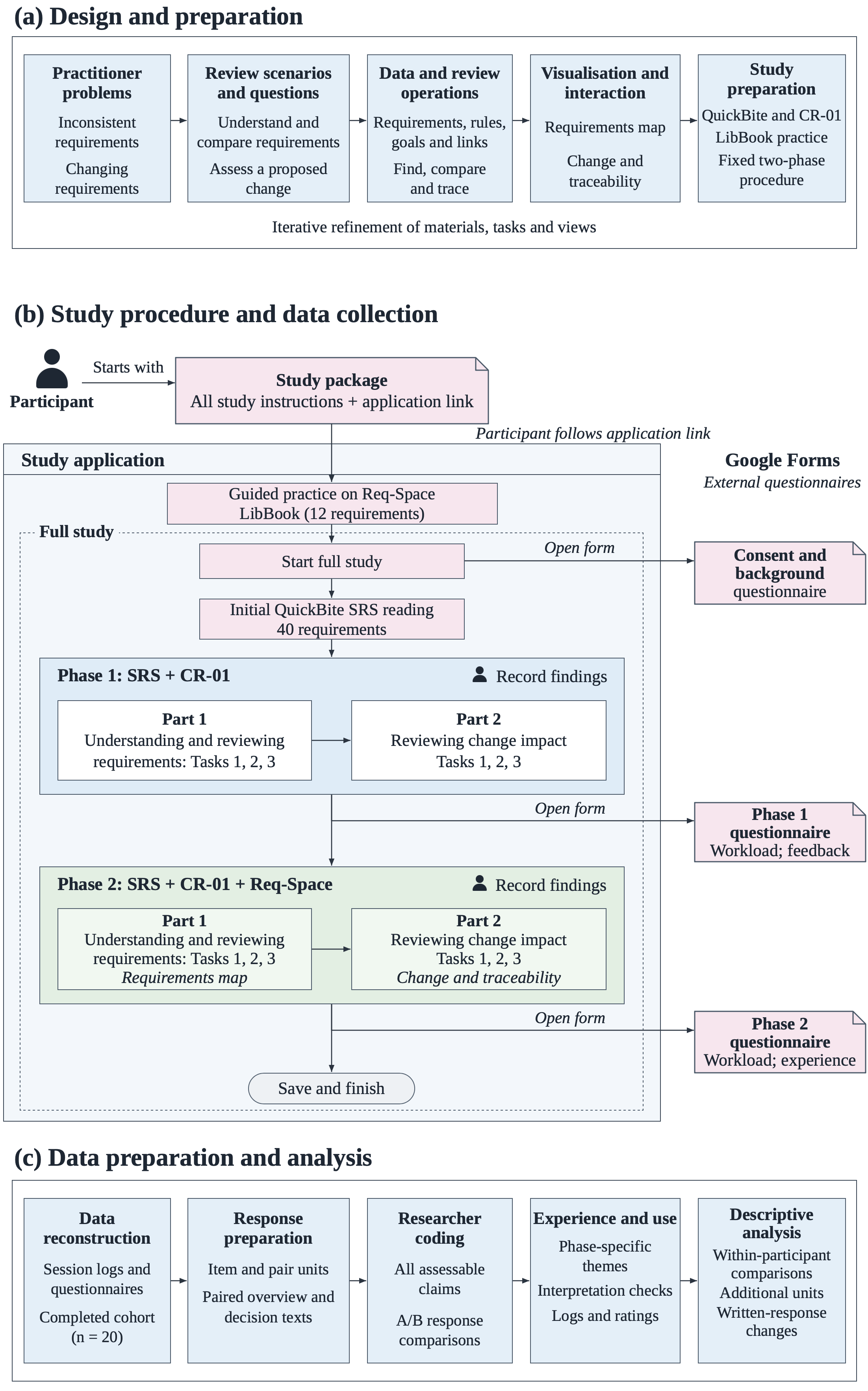}
    \caption{Overview of the study design: (a) design and preparation, (b) data collection, and (c) data preparation and analysis.}
    \label{fig:study-design}
\end{figure}

We conducted a fixed-order, two-phase study to examine additions and revisions to an initial requirements review. Participants first reviewed using documents alone, then revisited the same tasks with Req-Space alongside the documents. The design examines how review records develop during a subsequent supported review, rather than isolating the tool's causal effect. Figure~\ref{fig:study-design} summarises the design and preparation, data collection, and analysis.

\subsection{Review Tasks and Operationalisation}
\label{sec:design-rationale}

The six tasks cover inconsistency review (Part~1) and change-impact review (Part~2). We mapped each task to the evidence, review operations, visual support and recorded output shown in Table~\ref{tab:design-mapping} (Fig.~\ref{fig:study-design}(a)). Tasks combine focused investigation with open exploration; participants were not expected to examine every requirement or relationship. The table's questions summarise the administered instructions. Source documents were available in both phases, with visual support added in Phase~2.

\begin{table*}[t]
\caption{Review tasks and their operationalisation.
INC denotes inconsistency review; CI denotes change-impact review.}
\label{tab:design-mapping}
\centering
\setlength{\tabcolsep}{4pt}
\begin{tabularx}{\textwidth}{
    @{}
    >{\raggedright\arraybackslash}p{0.065\textwidth}
    >{\raggedright\arraybackslash}X
    >{\raggedright\arraybackslash}X
    >{\raggedright\arraybackslash}p{0.21\textwidth}
    >{\raggedright\arraybackslash}p{0.12\textwidth}
    @{}
}
\toprule
Task &
Reviewer question &
Evidence and review operations &
Phase~2 visual support &
Recorded output \\
\midrule

\multicolumn{5}{@{}l}{
\textbf{Part 1: Understanding and reviewing requirements}
} \\
\addlinespace

INC-T1 &
What does the specification cover? &
Browse requirements by feature, actor and source team; inspect details and system context. &
Requirements map; search and filters. &
Requirements overview \\
\addlinespace

INC-T2 &
Can requirements from Payments and Customer Support be satisfied together? &
Locate and compare relevant statements, considering source teams, rules and definitions. &
Search and filters; related-wording list; comparison tray. &
Findings \\
\addlinespace

INC-T3 &
Which other requirements warrant comparison or clarification? &
Select further requirements, compare statements and assess relationships using supporting SRS information. &
Requirements map; related-wording list; comparison tray. &
Findings \\

\midrule
\multicolumn{5}{@{}l}{
\textbf{Part 2: Reviewing a proposed change}
} \\
\addlinespace

CI-T1 &
What changes, and what consequences might follow? &
Compare current and proposed wording; examine the change rationale and related items. &
CR-01 panel; Change and traceability. &
Change impact overview \\
\addlinespace

CI-T2 &
Which items need revision, clarification, or no change? &
Follow recorded links and investigate other requirements, rules and goals to assess change effects. &
Change and traceability; search and filters; comparison tray. &
Findings \\
\addlinespace

CI-T3 &
Can the change proceed, or are revisions or clarification needed? &
Weigh the proposal, relevant SRS information and review findings; justify the decision and identify unresolved questions. &
CR-01 panel; Change and traceability; comparison tray. &
Decision rationale \\

\bottomrule
\end{tabularx}
\end{table*}

\subsection{Study Materials}
\label{sec:study-materials}

We developed the materials through iterative prompting with ChatGPT (GPT-5.6 Sol)~\cite{openai_chatgpt} and review by two authors to simplify wording and clarify scenarios.

The main review used QuickBite, a fictional food delivery system with 40 requirements across eight feature groups. Its SRS includes goals, business rules, scope and constraints, an order process, and a glossary. Requirements have stable identifiers, source teams, and recorded relationships to other requirements and supporting items. The accompanying change request, CR-01, proposes replacing mandatory sign-in with guest checkout using contact and delivery information. All participants reviewed the same SRS and change request in both phases.

Practice used LibBook, a separate fictional library-room booking system with 12 requirements, supporting information, and a proposed change. Worked examples and independent exploration introduced the visualisations, comparison controls, and finding-recording forms without exposing participants to QuickBite.

\subsection{Participants}
\label{sec:participants}

We approached more than 40 people through personal and professional contacts and a master's-level Requirements Engineering course at Tampere University, focusing recruitment on industry experience. Participation was voluntary; course recruits were offered a EUR~15 voucher. Twenty participants completed both phases, with an estimated time commitment of two hours including practice.
The sample included participants experienced in software/IT and requirements review (Table~\ref{tab:participant-background}). Requirements experience is reported separately from student status because these categories overlap.

\begin{table}[t]
\caption{Reported background of participants ($n=20$).
Roles and activities were multi-select.}
\label{tab:participant-background}
\centering
\footnotesize
\begin{tabular}{@{}lrr@{}}
\toprule
\multicolumn{3}{@{}l}{\textbf{Experience (years)}} \\
 & Software/IT & Requirements \\
\midrule
None                       & 1 & 1 \\
Less than 1                & 2 & 6 \\
1 to less than 3           & 6 & 4 \\
3 to less than 6           & 3 & 2 \\
6 to less than 10          & 5 & 4 \\
10 or more                 & 3 & 3 \\
\midrule
\multicolumn{2}{@{}l}{\textbf{Current roles}} & $n$ \\
\multicolumn{2}{@{}l}{Software developer / architect} & 9 \\
\multicolumn{2}{@{}l}{Student} & 9 \\
\multicolumn{2}{@{}l}{Researcher / teacher} & 6 \\
\multicolumn{2}{@{}l}{Other technical/management roles$^{a}$} & 5 \\
\midrule
\multicolumn{2}{@{}l}{\textbf{Previous RE activities}} & $n$ \\
\multicolumn{2}{@{}l}{Understanding requirements as a whole} & 18 \\
\multicolumn{2}{@{}l}{Reviewing for inconsistencies} & 12 \\
\multicolumn{2}{@{}l}{Checking effects of changes} & 13 \\
\multicolumn{2}{@{}l}{Tracking requirements relationships} & 10 \\
\bottomrule
\end{tabular}

\vspace{2pt}
\parbox{\columnwidth}{\footnotesize
$^{a}$Includes project/team leadership, testing/QA,
requirements analysis, product management, and
hardware design. Role categories may overlap.}
\end{table}

\subsection{Study Procedure and Data Collection}
\label{sec:data-collection}

Figure~\ref{fig:study-design}(b) summarises the procedure. Participants received instructions, completed guided practice with LibBook, then completed the consent and background questionnaire and read the QuickBite SRS.

In \textbf{Phase~1}, participants reviewed the requirements and proposed change using the SRS and CR-01. After the post-phase questionnaire, they repeated the same tasks in the same order in \textbf{Phase~2}, with Req-Space alongside the documents: the Requirements map for Part~1 and Change and traceability for Part~2. Questions remained unchanged, with no explicit request to improve earlier responses. Phase~1 responses were available as read-only references; Phase~2 responses were recorded separately. Participants worked at their own pace and completed a questionnaire after each phase.

\textbf{Review records} comprised three response types (Table~\ref{tab:design-mapping}): \textbf{overviews}, containing free-text observations without predefined topics; \textbf{findings}, identifying SRS items with a verdict, explanation and confidence rating; and \textbf{change decisions}, stating whether to accept CR-01 as written, accept it with revisions, or seek clarification, with reasons, questions and confidence. Confidence was rated low, medium or high. Phase~2 findings also included an optional field explaining what prompted them.

\textbf{Interaction logs} recorded navigation, document access, selections, feature use and response submissions or revisions, linked to participants, phases and tasks. They capture actions, not attention or reasoning.

\textbf{Questionnaires} collected background information on roles, experience, requirements activities, review practices and visualisation use. Both post-phase questionnaires included six NASA-TLX-adapted workload items~\cite{nasaTLX}: mental, physical and temporal demand, effort, frustration and perceived success. Items used a 0--10 scale and were analysed separately; higher success scores indicated greater success. They also collected five-category ease ratings for six review activities, perceived information sufficiency, and open-ended reflections on understanding requirements, reviewing inconsistencies and assessing change impacts.

The \textbf{Phase~2 questionnaire} additionally covered the helpfulness of six features, including non-use and inability-to-rate options; changes in findings or understanding; unclear or misleading elements; support needs; and potential use. Two multiple-choice questions assessed interpretation of map proximity and traceability links.

\subsection{Data Preparation}

We linked review records, interaction logs, and questionnaires by participant and phase, retaining the latest submissions from participants who completed both phases. Phase~1 and Phase~2 responses were retained separately for comparison.

For analysis, inconsistency findings were represented as judged item pairs and change-impact findings as individual judged items. Contextual references were excluded, and findings involving more than two judged items were split into all possible pairs. Single-item clarification findings were retained, while other incomplete consistency findings were flagged. Duplicate units within each participant, phase, and review part were counted once, retaining the latest verdict. We matched units across phases and classified Phase~2 units as additional (no Phase~1 match), repeated (unchanged verdict), or revised (changed verdict).

Overview responses and change-decision rationales were paired across phases by participant and task. Near-identical and referring-back responses were classified separately. The remaining pairs were presented in varying A/B order, with explicit phase cues masked. Decision categories remained attached to their rationales.

\subsection{Data Analysis}
\label{sec:data-analysis}

We analysed the three research questions using descriptive summaries and within-participant comparisons.

\paragraph{Review records (RQ1).}
We assessed distinct claims (referenced item or pair and verdict) against the QuickBite SRS and CR-01. Claims were labelled \emph{supported} if the verdict was defensible from the sources, including reasonable inferences or genuine ambiguity, and \emph{unsupported} otherwise. Identical claims within the same review part received the same label. Incomplete pairs were excluded from support percentages. We summarised additional, repeated and revised units, considering earlier support labels when interpreting revisions. Support assessment concerned the verdict, not the written explanation.

We separately compared overviews and decision rationales using a rubric covering correctness, relevance, errors and completeness in addressing the task. Pairs were labelled \emph{A better}, \emph{B better}, \emph{about the same} or \emph{not comparable}, then mapped back to their phases. Decision rationales were judged independently of the decision category, while changes in final decisions were counted separately.

One author coded 107 distinct assessable claims, 38 overview pairs and 17 decision-rationale pairs. Two coauthors jointly reviewed a subset of 26 claims, 20 overview pairs and 13 rationale pairs. The primary coder considered their assessments and revised three claim labels, seven overview comparisons and two rationale comparisons. Only the consolidated human labels were used in the results; independent inter-rater agreement was not estimated.

For descriptive comparison, we grouped participants by requirements experience: under one year, one to under six years, and six or more years. We compared individual supported-unit counts across these groups without testing an effect of experience.

\paragraph{Feature use and perceived support (RQ2).}
We counted participants with at least one corresponding Phase~2 interaction event and summarised feature helpfulness ratings separately. Negative ratings were combined as \emph{got in the way}, while non-use and inability to rate remained separate. Workload was compared using phase medians and within-participant counts of lower, unchanged and higher ratings. Higher scores indicated greater demand or effort, except for perceived success. Task-ease ratings were compared on their five-category scale, where higher scores indicated greater ease. Paired comparisons used available responses, with denominators reported.

\paragraph{Experiences and interpretation (RQ3).}

One author conducted a thematic analysis of paired open-ended responses concerning requirements understanding, inconsistency review and change effects. Responses were initially coded by phase, with AI assistance used to group descriptive codes into broader themes. The author refined the themes while retaining contrasting and mixed experiences, and a coauthor reviewed them. Comments on prior practices, unclear elements, further support needs and possible use supplemented the analysis. Quotations were checked against the original responses and included only with permission.

The two interpretation questions were scored as correct, incorrect or unsure, based on whether participants recognised map proximity as wording similarity and trace links as relationships recorded in the SRS.

%% file: sections/results.tex
\section{Results}
\label{sec:results}

The results cover the 20 participants who completed both review phases and the post-phase questionnaires. They include the range of experience reported in Table~\ref{tab:participant-background}. Phase~1 is the initial document-based review; Phase~2 is the subsequent review with the documents and Req-Space. The comparisons below describe this sequence, rather than an isolated effect of the visualisations.

\subsection{\textbf{RQ1: Additions and Revisions to Review Records}}
\label{sec:results-rq1}

\subsubsection{Additional findings and revised verdicts}

Most participants extended their initial review when revisiting the requirements with Req-Space. Seventeen of the 20 participants recorded at least one additional supported item or pair. Of the 135 units recorded in Phase~2, 70 were additional, 59 repeated earlier judgements, and six revised an existing verdict. Among the additions, 58 were supported by the source materials, 11 were unsupported, and one was not assessable. Thus, 58 of 69 assessable additions (84.1\%) were supported. These additions covered both review activities: 32 supported units from inconsistency review and 26 from change review.

The additions indicate broader recorded review coverage, rather than necessarily more defects. For example, P08 identified a conflict between two cancellation requirements: ORD-06 allows free cancellation until courier pickup, whereas PAY-04 requires full payment after restaurant acceptance. Both rules apply between acceptance and pickup, creating a conflict. Supported additions also include judgements of compatibility or that no change is needed. Across participants, the 58 supported additions extended the 158 supported units recorded in Phase~1 to a cumulative total of 216.

\begin{figure}[t]
    \centering
    \includegraphics[width=0.85\columnwidth]{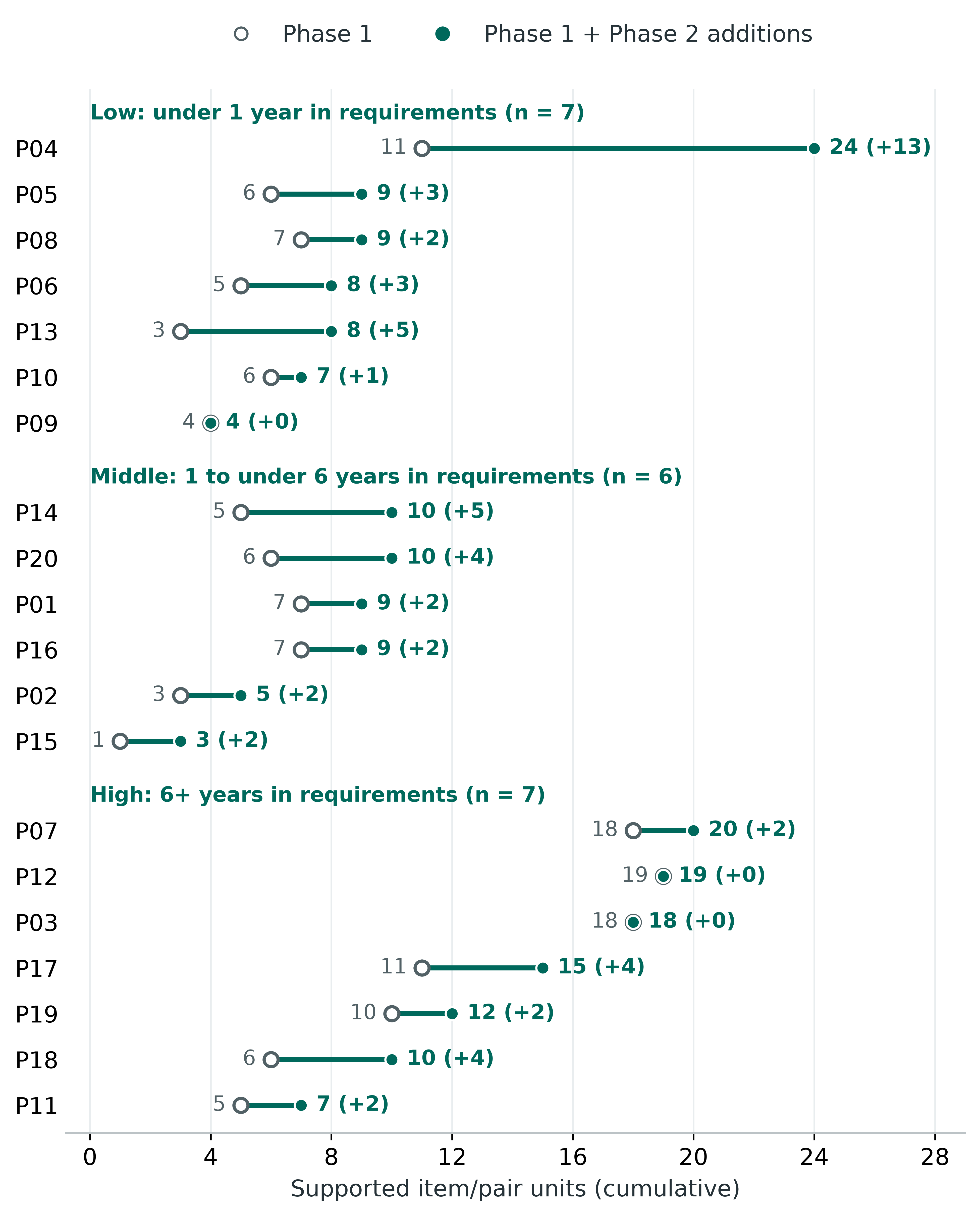}
    \caption{Supported item/pair units by requirements experience: under one year ($n=7$), one to under six years ($n=6$), and six or more years ($n=7$). Hollow circles indicate Phase~1 counts, and filled circles include supported Phase~2 additions. Parentheses show the number of additions. Rows are ordered by cumulative count within each group.}
    \label{fig:additional-units}
\end{figure}

Figure~\ref{fig:additional-units} shows the supported additions across participants with different levels of requirements experience. Most participants added supported findings across all three experience groups, but the number of additions varied considerably. For example, P04, with less than one year of experience, added 13 supported units, the highest among all participants. In contrast, two experienced participants (P03 and P12) recorded no additions, although they had already identified 18 and 19 supported units in Phase~1. These differences show that additional coverage was not limited to any particular experience group.

In contrast to the additional coverage, changes to previously recorded verdicts were uncommon. Only six Phase~2 units revised an earlier verdict. All six earlier verdicts had been supported; five revised verdicts remained supported, while one became unsupported. None corrected a previously unsupported verdict. Thus, the subsequent review primarily added to existing records rather than correcting earlier judgements.

\subsubsection{Overview responses and change decisions}

\begin{figure}[t]
    \centering
    \includegraphics[width=0.9\columnwidth]{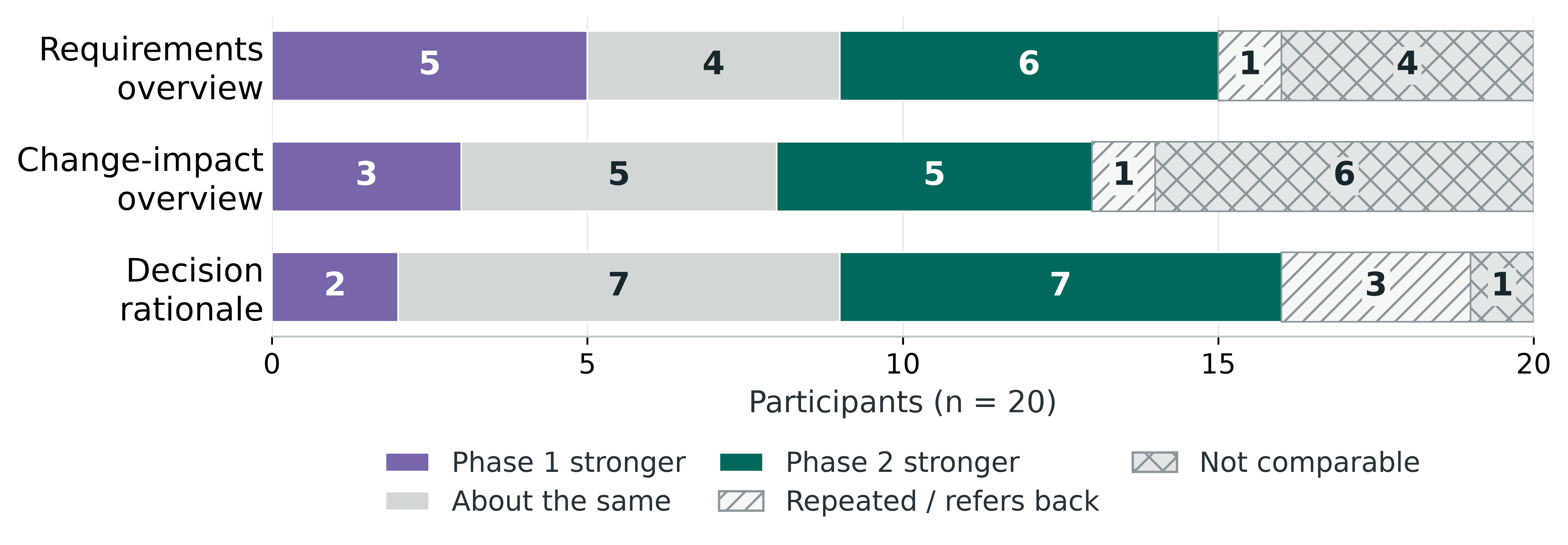}
    \caption{Comparison of written responses across
    phases ($n=20$ per task). Stronger responses were
    judged more accurate, relevant, and complete under
    the coding rubric. Repeated or referring-back
    responses and non-comparable pairs are shown
    separately.}
    \label{fig:written-comparisons}
\end{figure}

Phase~2 repeated the same review questions without asking participants to improve their earlier responses, which remained available for reference. We compared the written overviews and change-decision rationales for correctness, relevance, and how well they addressed the task. These comparisons assess the submitted responses, not participants' overall understanding. Figure~\ref{fig:written-comparisons} presents the results.

For requirements overviews, six responses were stronger in Phase~2, five in Phase~1, and four were about the same. For change-impact overviews, the corresponding counts were five, three, and five. Some stronger responses identified more specific relationships in the specification. For example, P20's Phase~2 change-impact overview identified a conflict between guest checkout and BR-1, which requires every order to belong to a registered account. This was judged stronger than the initial overview. Although the comparisons were mixed, some participants provided more informative and better-supported accounts during the subsequent review. The remaining pairs were repeated, referred back to earlier responses, or were not comparable.

Decision rationales showed a clearer difference: seven responses were stronger in Phase~2, two in Phase~1, and seven were about the same. Three were repeated or referred back, and one was not comparable. Notably, five of the seven participants with stronger Phase~2 rationales retained their original decision. This suggests that revisiting a change may help reviewers explain an existing judgement more clearly, without changing the judgement itself.

Only five participants changed their final decision on CR-01. Four moved from requiring clarification to accepting the change with revisions, while one moved in the opposite direction. The remaining 15 retained their decision, and none accepted the change as written in either phase.

Overall, supported additions were the most widespread change, but some participants also provided stronger overviews or decision rationales without being explicitly asked to improve their earlier explanations. Many responses remained similar or were not directly comparable. These results distinguish extending review coverage, changing decisions, and strengthening recorded justifications. The fixed study sequence prevents attributing these differences to Req-Space alone.

\subsection{\textbf{RQ2: Use and Perceived Support of Req-Space}}
\label{sec:results-rq2}

\subsubsection{Feature helpfulness and recorded interactions}

\begin{figure}[t]
    \centering
    \includegraphics[width=\columnwidth]{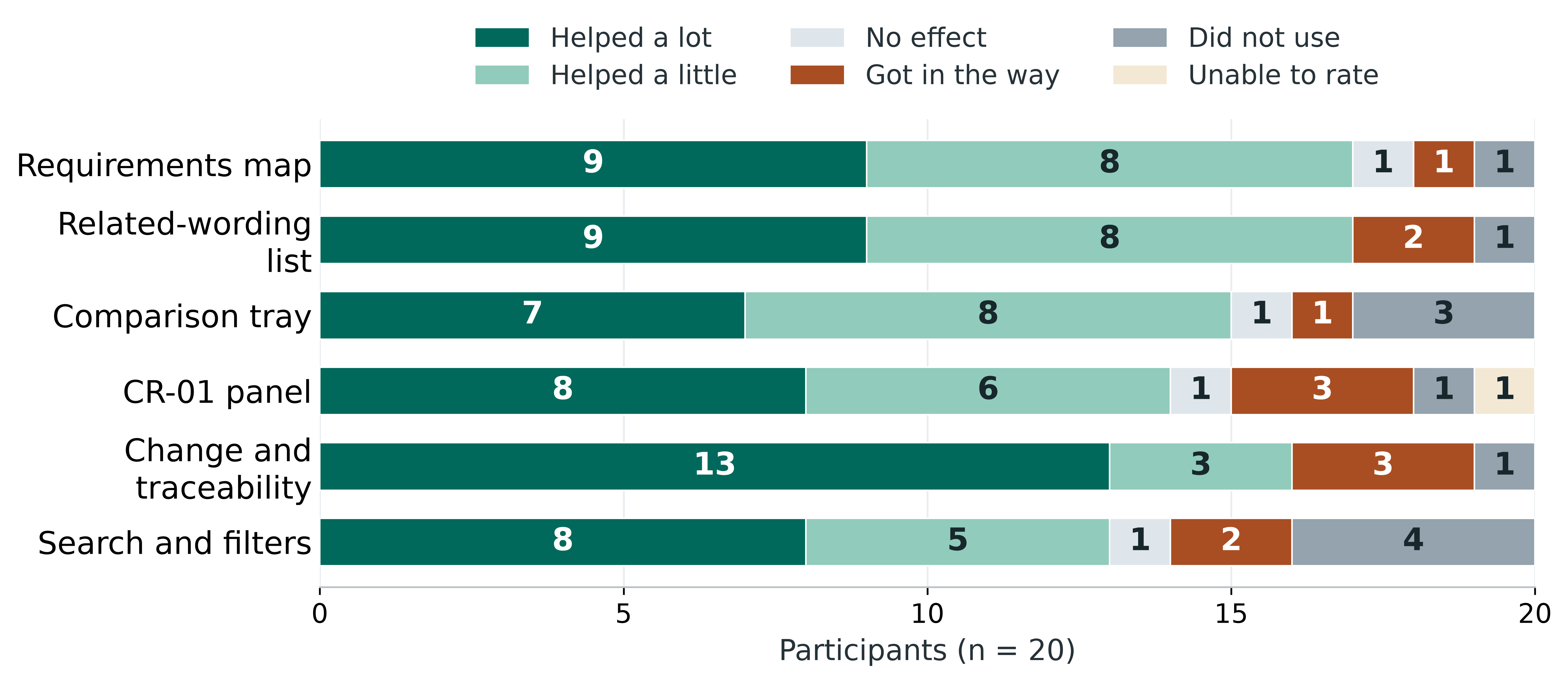}
    \caption{Perceived helpfulness of Req-Space features after Phase~2 ($n=20$). The two negative response levels are combined as \emph{got in the way}. Non-use and inability to rate are retained.}
    \label{fig:feature-helpfulness}
\end{figure}

Most participants rated the Req-Space features as helpful, although every feature received at least one negative rating (Fig.~\ref{fig:feature-helpfulness}). We examined these ratings alongside recorded interactions to understand how participants engaged with the features they evaluated.

The Requirements map and related-wording list each received 17 positive ratings. Map interactions were common: 16 participants selected requirements, 15 changed the colour mode, and 12 focused a legend category. In contrast, 13 participants opened the related-wording panel, and only five selected or compared an entry. Similarly, 15 participants rated the comparison tray positively, while ten added items to it.

Change and traceability received 16 positive ratings, including the most \emph{helped a lot} responses (13), although three participants reported that it got in the way. Thirteen participants expanded the trace diagram, while fewer filtered relationship types (4) or added items to the diagram (3). The CR-01 panel received 14 positive ratings, and nine participants interacted with it.

Search and filters showed a similar difference between ratings and recorded use. Thirteen participants rated them positively, although only one used text search and two changed requirement filters. Four participants reported not using these features.

Overall, positive helpfulness ratings were more common than some of the corresponding recorded interactions. However, the questionnaire grouped multiple controls under broader feature labels, and the logs captured specific actions rather than attention or reading. Participants could find displayed information helpful without interacting further. These aggregate comparisons therefore describe perceived helpfulness and recorded use separately, without establishing which interactions contributed to particular review findings.

\subsubsection{Workload and task difficulty}

Workload ratings showed mixed changes across phases (Table~\ref{tab:workload}). Eleven participants rated effort and hurried or rushed pace lower in Phase~2, although seven rated effort higher. Frustration and perceived success showed no clear directional pattern. Overall, the workload responses did not indicate a consistent reduction across participants or items.

\begin{table}[t]
\centering
\caption{Changes in workload ratings from Phase~1 to Phase~2 ($n=20$).}
\label{tab:workload}
\footnotesize
\setlength{\tabcolsep}{3pt}
\begin{tabular}{@{}lccc@{}}
\toprule
Item &
\shortstack{Lower score\\in Phase 2} &
\shortstack{Same score\\in both phases} &
\shortstack{Higher score\\in Phase 2} \\
\midrule
Mental demand       &  8 & 6 & 6 \\
Physical demand     &  7 & 7 & 6 \\
Hurried/rushed pace & 11 & 6 & 3 \\
Effort              & 11 & 2 & 7 \\
Frustration         &  6 & 7 & 7 \\
Perceived success   &  7 & 5 & 8 \\
\bottomrule
\end{tabular}
\begin{minipage}{\columnwidth}
\scriptsize
Values indicate the number of participants. Scores range from 0 to 10. Lower scores indicate less demand, effort, or frustration, except for perceived success, where higher scores indicate greater success.
\end{minipage}
\end{table}

Participants also rated how easy they found six review activities in each phase, using a five-point scale where higher scores indicated greater ease. In Phase~2, the median rating increased from 3 to 4 for finding related requirements, assessing inconsistencies, identifying change effects, and deciding whether to accept the change. The median remained at 4 for understanding the requirements as a whole and understanding CR-01. Each activity had 19--20 paired responses. Although participants rated four activities as easier in Phase~2, this may also reflect their familiarity with the materials from Phase~1, rather than the support provided by Req-Space alone.

\subsection{\textbf{RQ3: Review Experiences and Remaining Support Needs}}
\label{sec:results-rq3}

\subsubsection{Access to requirements and system context}

The 60 paired responses showed different benefits of the SRS and Req-Space. Participants valued the SRS structure, requirement IDs, and system context, but comparing information across sections required manual effort. In Phase~2, visual orientation was discussed by 16 participants, while 12 described easier navigation or comparison. P09 explained: \emph{``Instead of searching through long text, I could click on specific nodes to immediately highlight linked requirements and compare their rules side by side.''}

These benefits varied by task and participant. P19 found the actor and team views useful but considered the overall system easier to understand from the document. P04 similarly found related requirements easy to locate but reported that reading the SRS in Phase~1 had helped establish the system context. P18 was unsure whether Req-Space improved inconsistency review compared with documents, while P17 described the change view as reinforcing earlier assumptions rather than changing them.

\subsubsection{Limits of visual evidence}

Ten participants' Phase~2 responses concerned the continued need to read and interpret requirements. P03 noted that semantic similarity helped locate some potential conflicts, but differently worded requirements could still be functionally inconsistent. Similarly, participants found the change graph useful for following recorded dependencies, while questioning whether those links covered all relevant effects. P12 warned that the view could give a \emph{``false sense of completeness''} when potentially affected requirements had no recorded links. P04 also suspected missing or incorrect connections. These were reported concerns, not independently verified link errors.

To assess how participants interpreted the visualisations, we included two multiple-choice questions at the end of the Phase~2 questionnaire. One asked what proximity between requirements in the Requirements map represented, and the other asked what a link in Change and traceability indicated. Thirteen participants correctly identified map proximity as similarity in requirement wording; five interpreted it as a documented dependency, and two were unsure. Similarly, 13 correctly identified trace links as relationships recorded in the SRS, while four interpreted them as confirmed change effects and three believed they were inferred by Req-Space. Twelve participants answered both questions correctly. These responses indicate that some participants misunderstood what the visual relationships represented. Such misunderstandings may have influenced how they used the visualisations during review, although their effect on review outcomes was not assessed.

\subsubsection{Further support and possible use}

Eight participants said additional support would have helped, six said no, and six were unsure. Their requests concerned three areas. First, participants wanted clearer guidance for review judgements. P05 requested examples showing how to distinguish inconsistencies, compatible requirements, and possible change effects, while P11 found it unclear how to use the relationship graph during review. Second, they wanted better ways to check review coverage. P07 suggested indicating which requirements had not yet been accessed or reviewed, and P04 wanted to add relationships discovered during review rather than rely only on existing links. Third, participants wanted help acting on their findings. P11 suggested highlighting potential conflicts and generating proposed changes to downstream requirements, while P07 wanted impact comments generated from recorded findings. Other difficulties concerned interface navigation and guidance.

Fifteen participants said they could use Req-Space in work or studies, while five were unsure. However, P11 wanted support beyond browsing, including editing requirements and handling accepted changes. Overall, these responses indicate needs for tracking review coverage, interpreting findings, and acting on review decisions. They describe potential applications and further development needs, not demonstrated adoption in practice.

%% file: sections/discussion.tex
\section{Discussion}
\label{sec:discussion}

The findings suggest three practical lessons for requirements review. They derive from individual reviews of one fictional system; effects on team coordination and implementation remain to be evaluated.

\paragraph{Complement Existing Review Practices}

Visual support should fit specific review tasks and preserve access to source documents. P18 described quicker identification of relevant requirements through the dependency graph but was less certain about its value for inconsistency review. Keeping source text accessible lets reviewers move between visual relationships and the broader system context needed to assess them.
Evaluation should distinguish additional findings, changed decisions, stronger justifications, and confirmation of existing judgements. A review can strengthen the rationale without changing the decision. Whether stronger recorded rationales improve communication among developers, testers, and product owners requires collaborative evaluation.

\paragraph{Explain Relationship Meaning and Coverage}

Interfaces should clearly distinguish wording similarity from recorded specification relationships and explain that neither establishes a conflict or change effect. They should also make link origins, maintenance, and possible omissions explicit, helping reviewers recognise the limits of the displayed evidence.
Teams should use these views as starting points for investigation, including inspection of potentially relevant requirements beyond existing links. Guidance and worked examples could support correct interpretation, but their effectiveness requires evaluation.

\paragraph{Connect Findings to Software Change Processes}

Visual exploration could prepare discussions with product owners and implementation teams by bringing related requirements and supporting evidence together. For example, guest checkout raises questions about account-dependent behaviour, payments, and customer support. Resolving these questions requires domain knowledge, source inspection, and stakeholder agreement.
To carry findings into implementation planning and testing, support should extend to tracking unreviewed requirements and linking findings to change requests, responsible teams, unresolved questions, and subsequent revisions. These workflows were not implemented or evaluated in Req-Space.
Automated assistance could generate impact comments and propose downstream requirement revisions, with suggestions linked to source evidence for reviewers to inspect, correct, or reject. Evaluation should examine missed issues, false alarms, and verification effort before claiming improved review quality or reduced workload.

\emph{Sustainability considerations.}
Req-Space reuses precomputed embeddings and layouts, with no language-model calls during review sessions. Emissions from preparation, hosting, and study analysis were not evaluated.

%% file: sections/threats_to_validity.tex
\section{Threats to Validity}
\label{sec:threats-to-validity}

We consider four validity threats~\cite{wohlin2012experimentation}.

\paragraph{Construct Validity}
Item/pair counts measure recorded review coverage, not independent defects or completeness. Supported verdicts include compatibility and no-change judgements, with source support assessed separately from explanations. Written comparisons capture expressed reasoning, ratings capture perceived helpfulness and workload, and logs record actions rather than attention or reasoning.

The two-dimensional t-SNE map cannot preserve all semantic relationships~\cite{vanDerMaaten2008tsne}, and proximity or related wording does not establish inconsistency. Trace links represent recorded SRS relationships, may be incomplete, and do not confirm change impacts~\cite{clelandhuang2014traceability}. Both views require source inspection and human judgement; the two interpretation questions only partially assess participants' understanding.

\paragraph{Internal Validity}
The fixed task order, prior practice, and availability of Phase~1 responses during Phase~2 introduce possible familiarity, recall, additional-review-time, and fatigue effects. Without a repeated document-only condition, these cannot be separated from Req-Space's contribution. Phase~2 therefore represents a subsequent supported review, not an independent treatment. One participant disclosed AI-assistant use; exclusion left additional-unit counts unchanged, but other outcomes were not checked.

\paragraph{External Validity}
The 20 volunteers, recruited through personal contacts and a university course, had varied software development and requirements-engineering experience but do not represent a defined industrial practitioner population. They individually reviewed one fictional specification with 40 requirements and one proposed change, using predefined relationships and fixed layouts.

Differences in terminology, requirements structure, traceability completeness, and workflows limit generalisation to larger or evolving industrial projects. The study concerns human-led evidence exploration, not automated inconsistency detection, exhaustive change-impact analysis, or formal verification. Scalability, data maintenance, collaborative industrial use, and applicability beyond software requirements review remain unexamined.

\paragraph{Conclusion Validity}
With 20 participants, individual responses can influence distributions, and units from the same participant are not independent. Comparisons are therefore descriptive, not causal. The rubric evolved during analysis, and coding agreement was not measured. Researcher judgement and AI-assisted theme grouping may have influenced interpretation. A/B order variation, masked phase cues, source-checked quotations and record-mapping checks reduced but did not eliminate interpretive bias.